\PassOptionsToPackage{hyphens}{url} 
\documentclass[%
 aip,
 amsmath,amssymb,
 reprint,%
]{revtex4-1}
\usepackage{graphicx}
\usepackage{dcolumn}
\usepackage{bm}
\usepackage[utf8]{inputenc}
\usepackage[T1]{fontenc}
\usepackage{mathptmx,comment}
\usepackage{amsmath, amssymb, mathtools, bm, etoolbox, comment}
\usepackage{setspace}
\usepackage{mathrsfs}
\usepackage{stmaryrd}
\usepackage{amsthm}
\usepackage{amsfonts}
\usepackage{graphicx}
\usepackage[colorinlistoftodos]{todonotes}
\usepackage{tikz-cd}   
\usepackage{booktabs}       
\usepackage{nicefrac}       
\usepackage{microtype}      
\usepackage[utf8]{inputenc} 
\usepackage[T1]{fontenc}    
\usepackage{hyperref}       
\usepackage{url}            
\usepackage{lingmacros}
\usepackage{tree-dvips} 
\usepackage{nccmath}
\usepackage{bbm}

\DeclareMathOperator*{\loggrad}{log-grad}

\usepackage[utf8]{inputenc} 
\usepackage[noend]{algpseudocode}
\usepackage{algorithm}
\usepackage{caption}
\usepackage{subcaption}
\usepackage{color}
\usepackage{enumitem}

\usepackage{amsmath}

\begin{document}

\preprint{AIP/123-QED} 
\title[COVID-19 in the United States: Trajectories and second surge behavior]{COVID-19 in the United States: Trajectories and second surge behavior}

\author{Nick James}
\affiliation{ 
School of Mathematics and Statistics, University of Sydney, NSW, 2006, Australia}%
\author{Max Menzies}
\email{max.menzies@alumni.harvard.edu}
\affiliation{%
Yau Mathematical Sciences Center, Tsinghua University, Beijing, 100084, China}%

\date{4 August 2020}
\begin{abstract}

This paper introduces a mathematical framework for determining second surge behavior of COVID-19 cases in the United States. Within this framework, a flexible algorithmic approach selects a set of turning points for each state, computes distances between them, and determines whether each state is in (or over) a first or second surge. Then, appropriate distances between normalized time series are used to further analyze the relationships between case trajectories on a month-by-month basis. Our algorithm shows that 31 states are experiencing second surges, while 4 of the 10 largest states are still in their first surge, with case counts that have never decreased. This analysis can aid in highlighting the most and least successful state responses to COVID-19.

\end{abstract}

\maketitle

\begin{quotation}

The United States (US) has been severely impacted by COVID-19, and leads the world in both case and death counts. \cite{worldindata2020} Individual states have largely determined their own response to the pandemic, \cite{Haffajee2020} seeking to protect citizens' lives while mitigating economic consequences. Following lockdowns and business closures in March and April, all 50 states reduced their restrictions. \cite{wapo_allreopen} Since then, the US has experienced a severe rise in new cases, with a public debate on whether or not to term this a "second surge." \cite{atlantic_secondsurge,Hopkinsmedicinearticle} A careful identification of most and least successful states is therefore of great relevance to a response to the ongoing threat of COVID-19. This paper has two goals: to develop a mathematical framework that defines a second surge and identifies the states that are experiencing one; and to compare the trajectories of new case counts as a means of determining effective pandemic responses. 

\end{quotation}

\section{Introduction}
\label{sec:intro}
Understanding the trajectory of COVID-19 case counts assists governments in responding to the impact of the pandemic. Given the highly infectious nature of the disease, an increasing number of new daily cases may overwhelm the healthcare system and prompt further restrictions on businesses and social gatherings. Conversely, a decreasing number of new cases is a good sign, but should be carefully monitored. Thus, the \emph{turning points} in the new case counts are crucial to identify. Given the randomness of human behaviour, however, turning points are difficult to predict and predictive modelling of COVID-19 requires frequent observation of real-time data. This paper focuses on a retrospective analysis of case trajectories in different US states to determine which public policy responses were most and least effective. While we focus on the US, our methodology may be applied more broadly to understand the impact of public policies on countries with similar federative structures such as Brazil and India. These three countries have the highest COVID-19 case counts in the world,\cite{worldindata2020} with government responses differing between states and yielding differing results. \cite{DaSilvamed,india}

For this goal, we use existing and new techniques from \emph{time series analysis}. Time series analysis has been widely applied to epidemiology, \cite{Hethcote2000,Chowell2016} including COVID-19. \cite{Manchein2020, Machado2020,James2020_chaos} Existing methods of time series analysis are diverse, including power-law models, \cite{Vazquez2006} and nonparametric  methods such as distance analysis, \cite{Moeckel1997} distance correlation \cite{Szkely2007,Mendes2018,Mendes2019} and network models. \cite{Shang2020} In this paper, we aim to develop a new mathematical framework of identification and comparison of \emph{turning points} of time series to study the spread of COVID-19 in the US. Such turning points classify the behavior of states' trajectories throughout the pandemic as being in (or over) their first surge or second surge.

In addition to the aforementioned state-by-state determination of turning points, this paper uses a new application of semi-metrics to measure distance between the states' behaviors, and performs clustering based on this. The paper implements \emph{hierarchical clustering},\cite{Ward1963,Szekely2005}
which has previously been used in various epidemiological applications. These include inflammatory diseases, \cite{Madore2007} airborne diseases, \cite{Kretzschmar2009} Alzheimer's disease,  \cite{Alashwal2019}  Ebola, \cite{Muradi2015} SARS, \cite{Rizzi2010} and COVID-19. \cite{Machado2020}

The paper is structured as follows: in each of the proceeding two sections, we introduce portions of our methodology and then present our results. Section \ref{sec:ChangePointmodel} describes our framework for identifying turning points, determining which states are in a second surge, and clustering based on similar behavior. Section \ref{sec:DynamicTrajectoryModelling} analyzes the trajectories of COVID-19 counts in each state on a month-by-month basis. Section \ref{sec:conclusion} summarizes the results and new findings regarding the spread of COVID-19 in the United States. In Appendix \ref{Brazil}, we briefly apply our analysis to the Brazilian states to demonstrate the generality of our method.

\section{Second surge analysis}
\label{sec:ChangePointmodel}

In this section, we develop a mathematical framework and procedure to determine whether a state has experienced a second surge. Through the careful selection of turning points, we formulate a definition applicable to an individual time series, and develop a method for comparing differing surge behaviors among a collection of time series.

\subsection{Second surge methodology: determination of turning points}
\label{sec:secondsurgemethod}
Let $x_i(t) \in \mathbb{R}$ be a collection of real-valued time series over a common time interval $t=1,\dots,T, i=1,\dots,n$. In this paper, the analyzed time series are the daily counts of new cases in the 50 US states and the District of Columbia (D.C.), ordered alphabetically, $i=1,\dots,51$. Our data spans 01/21/2020 to 07/31/2020, a period of $T=193$ days across $n=51$ regions. 

There are several irregular features in the data set, including lower case counts on the weekends, some negative daily counts due to adjustments of previous figures, and general noise. In addition, there are small disparities between different data sources. In order to isolate the signal in a data set, and between different data sets, 
we first apply a \emph{Savitzky-Golay filter} \cite{Savitzky1964} to the counts to produce a collection of \emph{smoothed time series} $\hat{x}_i(t)$,  $t=1,\dots,T, i=1,\dots,n.$ This combines a moving average calculation with polynomial smoothing. Through its moving average computations, it largely eliminates all negative counts, except when there are very few cases. In these instances, we replace any negative smoothed count with a zero. For the remainder of the paper, we analyze these smoothed cases $\hat{x}_i(t)$. Due to the smoothing process, $\hat{x}_i(t)\in\mathbb{R}_{\geq 0}$ are not necessarily integers, but are all non-negative.

Our identification of turning points and second surge behavior of a smoothed time series $\hat{x}(t)$ proceeds in two steps. First, we identify a sequence of potential local maxima or \emph{peaks}, and local minima or \emph{troughs}. Second, we appropriately refine this sequence according to chosen conditions. The final sets $P$ and $T$ of peaks and troughs, respectively, determine whether the time series is said to be in (or over) its first or second surge. It will be essential that $P$ and $T$ are non-empty.

For the first step, the basic idea is to designate a time $t_0$ a peak or trough, respectively, if 
\begin{align}
\label{baddefnpeak}
\hat{x}(t_0)&=\max\{\hat{x}(t): \max(1,t_0 - l) \leq t \leq \min(t_0 + l,T)\}\\
\label{baddefntrough}\hat{x}(t_0)&=\min\{\hat{x}(t): \max(1,t_0 - l) \leq t \leq \min(t_0 + l,T)\}
\end{align}
for a parameter $l<\frac{T}{2}$, the length over which we look. In this paper, we select $l=17$ to account for the 14-day incubation period of the virus \cite{incubation2020} and reduced testing on weekends. These naive definitions (\ref{baddefnpeak}, \ref{baddefntrough}) have two flaws: first, equal values of $\hat{x}(t)$ may determine consecutive values of $t$ as peaks or troughs when only one should be counted. More subtly, it is possible that two troughs may be detected at two points that are far apart, with no peak between them, when the time series has been largely monotonic between the two. For example, in Figure \ref{fig:MississippiTS}, troughs are naively detected at $t_0=1$ and $126$, corresponding to the start of the time series and 05/26/2020, respectively. What follows is a method to exclude the latter.

We implement variants of (\ref{baddefnpeak}, \ref{baddefntrough}) by sequentially examining the values of $\hat{x}(t)$. The first peak or trough is assigned at the first value of $t_0$ such that (\ref{baddefnpeak}) or (\ref{baddefntrough}) holds, respectively. For each US state, this is an initial trough at $t_0=1$ corresponding to zero cases there after smoothing. Having determined a peak at $t_0$, we search in the period $t>t_0$ for one of two elements: if we find a trough at $t_1>t_0$ according to (\ref{baddefntrough}) we add it to the set of troughs and proceed from $t_1$ as normal. If we find a peak at $t_1>t_0$ according to (\ref{baddefnpeak}) such that  $\hat{x}(t_0)\geq \hat{x}(t_1)$, we ignore this lesser peak as redundant; if we find a peak at $t_1>t_0$ according to (\ref{baddefnpeak}) such that  $\hat{x}(t_0) < \hat{x}(t_1)$, we remove the peak $t_0$ and replace it with $t_1$ and continue from there. An analogous process applies from a trough at $t_0$. With this process, we generate an alternating sequence of troughs and peaks, starting with a trough at $t_0=1$. Every time series is assigned at least one peak and trough at its global maximum and minimum, respectively. If the global maximum is not unique, a peak is assigned at the first maximum. This concludes the first step.

So far, every time series in our collection is assigned an alternating sequence of peaks and troughs beginning with a trough at $t_0=1$. One could naively define a state as being in its second surge if its sequence so far is TPTP. However, some detected peaks and troughs are immaterial and should be excluded. We describe a flexible approach to excluding trivial peaks or troughs, in which we apply two conditions to do so. 

Let $t_1<t_3$ be two peaks, necessarily separated by a trough. We select a parameter $\delta$, and if the \emph{peak ratio}, defined as $\frac{\hat{x}(t_3)}{\hat{x}(t_1)}<\delta$, we remove the peak $t_3$. If two consecutive troughs $t_2,t_4$ remain, we remove $t_2$ if $\hat{x}(t_2)>\hat{x}(t_4)$, otherwise remove $t_4$. That is, if the second peak has size less than $\delta$ of the first peak, we remove it, deciding not to term it a second surge.

Finally, we define the \emph{log-gradient} between times $t_1<t_2$ as 
\begin{align}
\label{loggrad}
   \loggrad(t_1,t_2)=\frac{\log \hat{x}(t_2) - \log \hat{x}(t_1)}{t_2-t_1}.
\end{align}
The numerator coincides with $\log(\frac{\hat{x}(t_2)}{\hat{x}(t_1)})$, and is a more appropriate substitution for the "rate of increase" given by $\frac{\hat{x}(t_2)}{\hat{x}(t_1)} -1$. Indeed, a "rate of increase" is asymmetrically bounded between $(-1,\infty)$ while the logarithmic rate is bounded between $(-\infty,\infty)$. The $\loggrad$ function measures the average rate of logarithmic increase over the period $[t_1,t_2]$. Now let $t_1,t_2$ be an adjacent peak and trough. We select a parameter $\epsilon=0.01$;  if
\begin{align}
    |\loggrad(t_1,t_2)|<\epsilon,
\end{align}
that is, the average logarithmic increase or decrease is well-defined and less than 1\%, we remove both $t_1$ and $t_2$ from our sets of peaks and troughs. For example, this step removes a peak and trough from Figure \ref{fig:GeorgiaTS}, where the local maximum  at 04/17/2020 is immaterial. This condition always preserves the trough at $t_0=1$, where $\hat{x}(t_0)=0$, and the peak at the global maximum. This concludes the selection of $P$ and $T.$

To quantify distance between time series' turning points, we apply the semi-metrics of Ref. \onlinecite{James2020_nsm} (with $p=1$). Given two non-empty finite sets $S_1,S_2$, this is defined as 
\begin{align}
\label{eq:MJ}
    D({S_1},{S_2}) = \frac{1}{2} \left(\frac{\sum_{b\in S_2} d(b,S_1)}{|S_2|} + \frac{\sum_{a \in {S_1}} d(a,S_2)}{|S_1|} \right),
\end{align}
where $d(b,S_1)$ is the minimal distance from $b \in S_2$ to the set $S_1$. The values $d(S_1,S_2)$ are symmetric, non-negative, and zero if and only if $S_1=S_2.$ Then, we define the $n \times n$ \emph{surge behavior matrix} between turning point sets by
\begin{align}
D_{ij}^{TP} = D(P_i,P_j) + D(T_i,T_j).
\end{align}
Then, $D_{ij}^{TP}=0$ if and only if time series $\hat{x}_i(t)$ and $\hat{x}_j(t)$ have equal sets of peaks and troughs, hence identical surge behavior. These procedures are presented in an algorithmic format in Appendix \ref{algorithm}.

\subsection{Second surge analysis results}
\label{secondsurgeresults}
\begin{figure*}
    \centering
            \begin{subfigure}[b]{0.33\textwidth}
        \includegraphics[width=\textwidth]{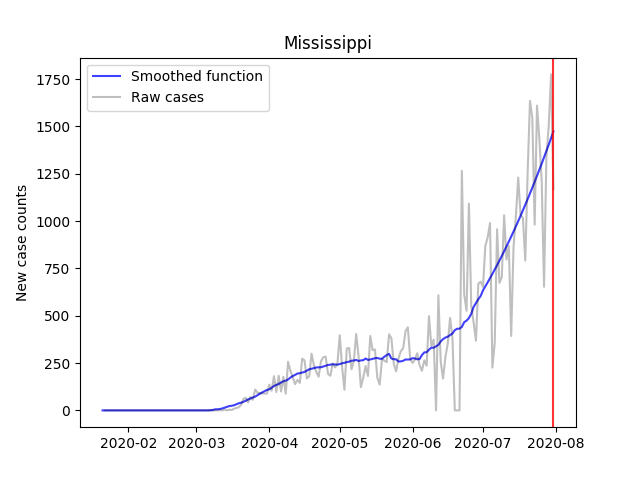}
        \caption{}
        \label{fig:MississippiTS}
    \end{subfigure} 
        \begin{subfigure}[b]{0.33\textwidth}
        \includegraphics[width=\textwidth]{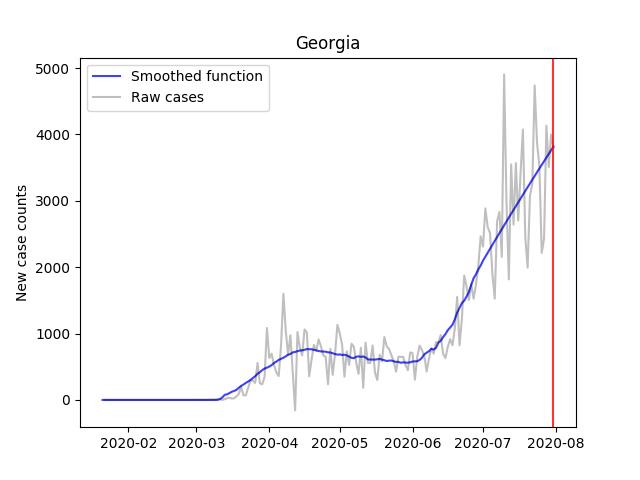}
        \caption{}
        \label{fig:GeorgiaTS}
    \end{subfigure}
        \begin{subfigure}[b]{0.33\textwidth}
        \includegraphics[width=\textwidth]{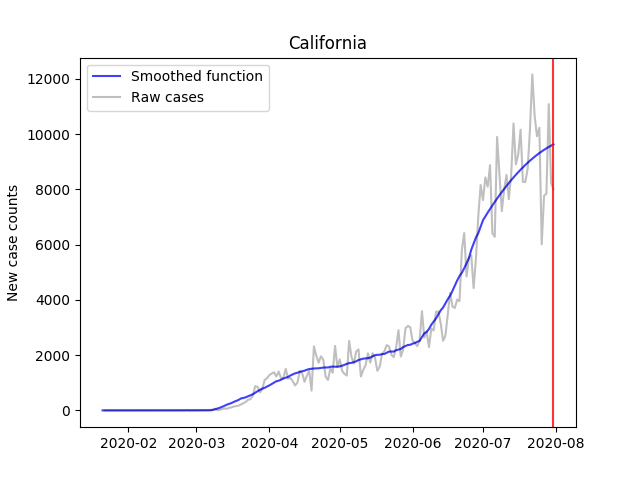}
        \caption{}
        \label{fig:CaliforniaTS}
    \end{subfigure}
    \begin{subfigure}[b]{0.33\textwidth}
        \includegraphics[width=\textwidth]{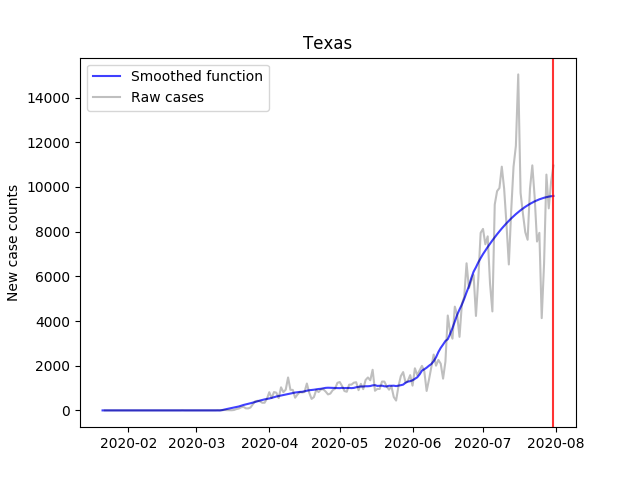}
        \caption{}
        \label{fig:TexasTS}
    \end{subfigure}
        \begin{subfigure}[b]{0.33\textwidth}
        \includegraphics[width=\textwidth]{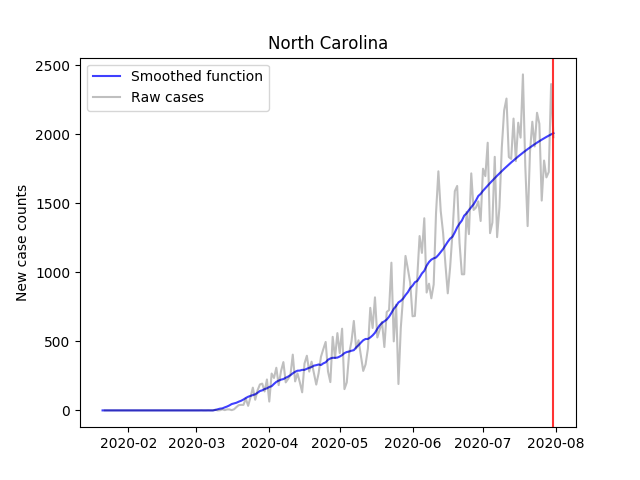}
        \caption{}
        \label{fig:NorthCarolinaTS}
    \end{subfigure}     
    \begin{subfigure}[b]{0.33\textwidth}
        \includegraphics[width=\textwidth]{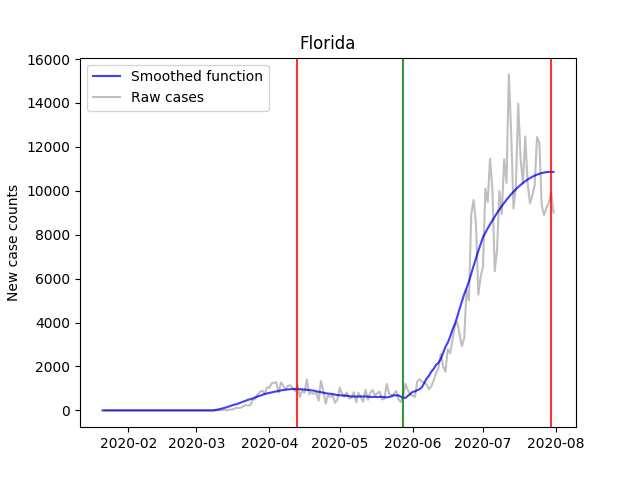}
        \caption{}
        \label{fig:FloridaTS}
    \end{subfigure}
       \begin{subfigure}[b]{0.33\textwidth}
        \includegraphics[width=\textwidth]{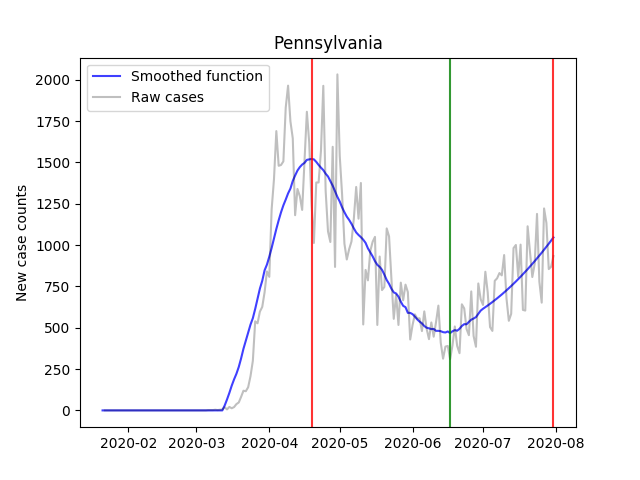}
        \caption{}
        \label{fig:PennsylvaniaTS}
    \end{subfigure} 
        \begin{subfigure}[b]{0.33\textwidth}
        \includegraphics[width=\textwidth]{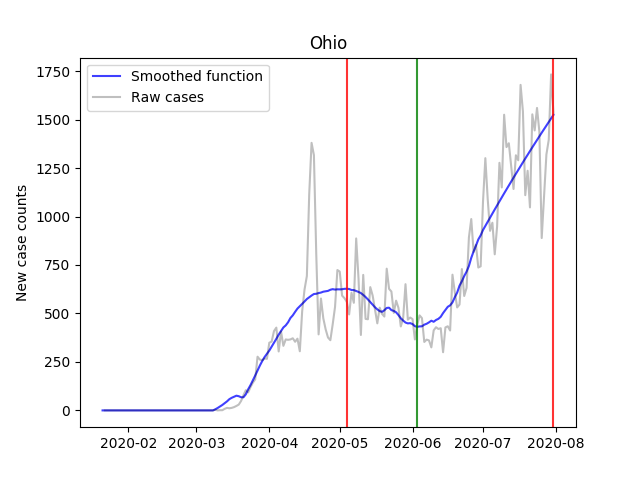}
        \caption{}
        \label{fig:OhioTS}
    \end{subfigure} 
\begin{subfigure}[b]{0.33\textwidth}
        \includegraphics[width=\textwidth]{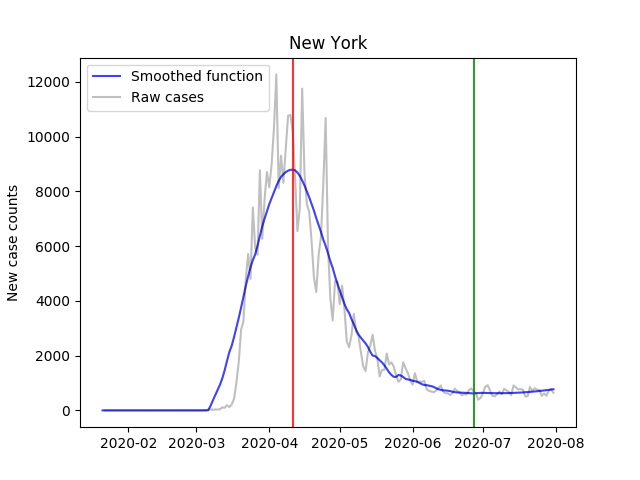}
        \caption{}
        \label{fig:NewYorkTS}
    \end{subfigure}     
\begin{subfigure}[b]{0.33\textwidth}
        \includegraphics[width=\textwidth]{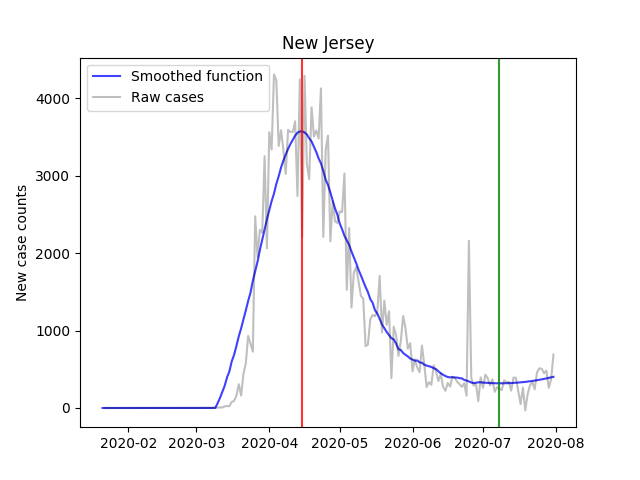}
        \caption{}
        \label{fig:NewJerseyTS}
    \end{subfigure} 
        \begin{subfigure}[b]{0.33\textwidth}
        \includegraphics[width=\textwidth]{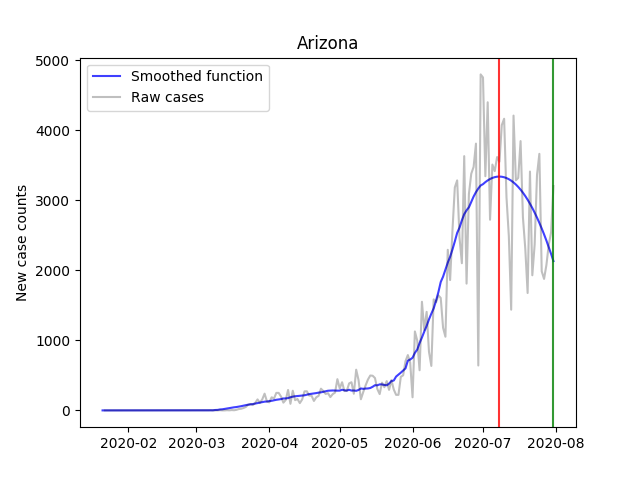}
        \caption{}
        \label{fig:ArizonaTS}
    \end{subfigure}
    \begin{subfigure}[b]{0.33\textwidth}
        \includegraphics[width=\textwidth]{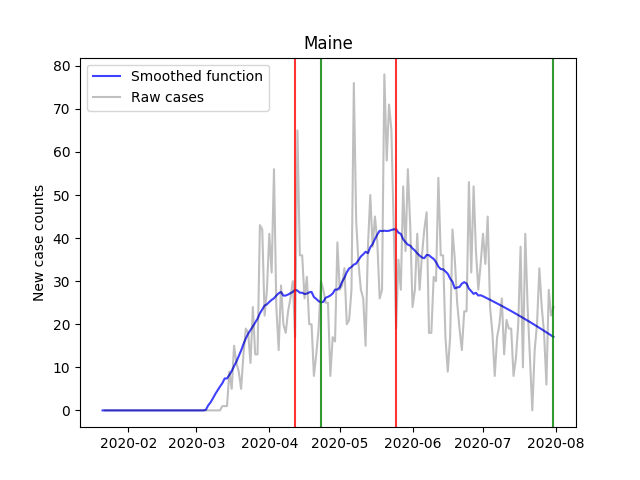}
        \caption{}
        \label{fig:MaineTS}
    \end{subfigure} 
    \caption{Smoothed time series and identified turning points for various states: (a) Mississippi (b) Georgia (c) California (d) Texas and (e) North Carolina are assigned sequence TP and determined to be in their first surge. (f) Florida (g) Pennsylvania and (h) Ohio are determined to be in their second surges, with sequence TPTP. (i) New York and (j) New Jersey are assigned sequence TPT and determined to have concluded their first surge and flattened the curve. (k) Arizona and (l) Maine are assigned TPT and TPTPT with final trough at the end of the period and determined to be declining from their first and second surges, respectively.}
    \label{fig:section2TSplots}
\end{figure*}

\begin{figure*}
    \centering
    \begin{subfigure}[b]{0.735\textwidth}
        \includegraphics[width=\textwidth]{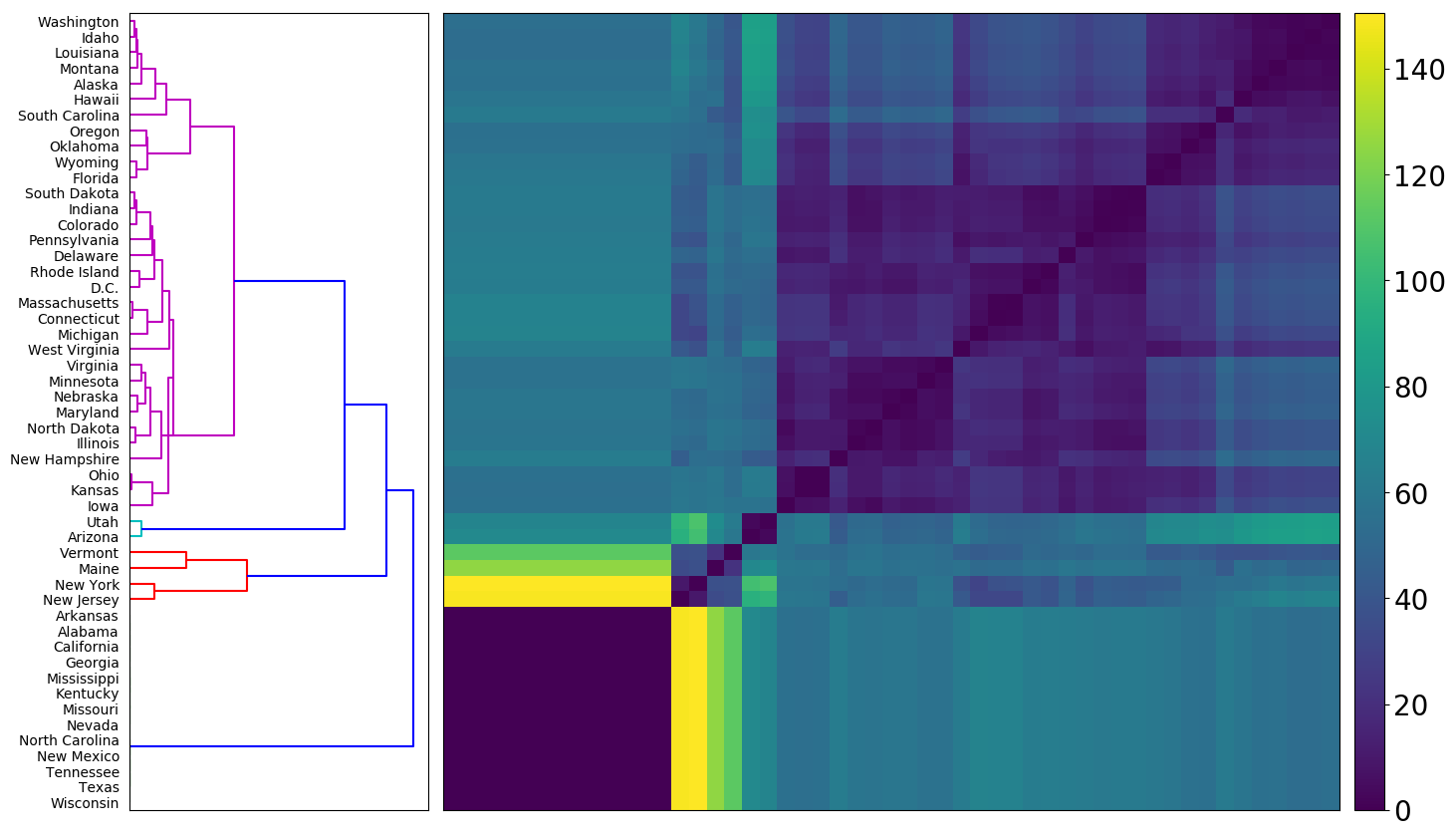}
    \end{subfigure}
    \caption{\emph{Surge behavior matrix}, defined in Section \ref{sec:ChangePointmodel}, measures distance between sets of turning points in new case trajectories. Five primary (sub)clusters of time series are identified with the following behaviors:  13 states in their first surge, 2 states that have completed their first surge and flattened the curve, 2 states coming down from their first surge, 31 states plus D.C. that are beyond their first surge and are now experiencing a second surge and 2 states coming down from their second surge.}
    \label{fig:MJ_Dendrogram}
\end{figure*}

Our methodology assigns one of four possible sequence types to each state. Thirteen states, including Georgia, California, Texas, and North Carolina (Figures \ref{fig:GeorgiaTS}, \ref{fig:CaliforniaTS}, \ref{fig:TexasTS}, and \ref{fig:NorthCarolinaTS}, respectively) are assigned the sequence TP (that is, one trough, then one peak) and deemed to be in their first surge. All 13 of these have their unique peak and global maximum on the final day, and form a cluster of identical similarity in Figure \ref{fig:MJ_Dendrogram}, where we implement hierarchical clustering on $D^{TP}$. Identical results are obtained with any value $\delta \in [0.1, 0.2]$, so we select $\delta=0.2$. That is, we exclude any second surge that is less than a fifth of the first. Then, 31 states plus D.C. are assigned TPTP - we deem these to be in their second surge. The three largest of these second surge states are Florida, Pennsylvania and Ohio, displayed in Figures \ref{fig:FloridaTS}, \ref{fig:PennsylvaniaTS} and \ref{fig:OhioTS}, respectively. Of these, all but Florida and South Carolina have a peak on the final day, with 19 exhibiting their global max on that day. These second surge states form the majority cluster in Figure \ref{fig:MJ_Dendrogram}. Four states are assigned the sequence TPT, of which New York and New Jersey (\ref{fig:NewYorkTS} and \ref{fig:NewJerseyTS}) have a local max removed due to a peak ratio less than 0.2. Their curves have been flattened and the first surge is completely over. Arizona (\ref{fig:ArizonaTS}) and Utah are also assigned TPT, with their latter trough on the final day, indicating they are still coming down from a first surge. Finally, Maine (\ref{fig:MaineTS}) and Vermont are assigned the sequence TPTPT. For Maine, this final trough is at the end of the period, indicating it is still coming down from its second surge, while Vermont's final trough is before the end, indicating it has flattened the curve on its second surge. Hierarchical clustering on $D^{TP}$ distinguishes all these behaviors into separate clusters.

\section{Dynamic trajectory modelling}
\label{sec:DynamicTrajectoryModelling}

In this section, we further analyze the new case counts in the 50 states and D.C., examining the trajectories of smoothed case counts on a month-by-month basis. Restricting these smoothed counts to a particular month gives a sequence $\mathbf{f}_i=({f}_i(1), {f}_i(2), \dots, {f}_i(m))) \in \mathbb{R}^m$, where $m \in \{29,30,31\}$ is the number of days in that month, and $i=1,\dots, 51$.

Let $||\mathbf{f}_i||=\sum_{t=1}^m |{f}_i(t)|$ be the $L^1$ norm of the vector $\mathbf{f}_i$. As all ${f}_i(t)$ are non-negative, this counts the total number of new cases (up to smoothing) observed in a month, and is non-zero for every state and month after March. Thus, we may define $\mathbf{g}_i= \frac{\mathbf{f}_i}{||\mathbf{f}_i||}$. The vectors $\mathbf{g}_i$ reflect relative changes of new case counts within a month. For example, a state whose new cases in a month differ between 1000 and 1100 will present a relatively flat normalized trajectory; whereas a state whose new cases in a month rise from 0 to 100 will present a more steeply increasing normalized trajectory, as a reflection of the relative change. We define \emph{trajectory distance matrices} $D_{ij}=||\mathbf{g}_i-\mathbf{g}_j||$ that measure distance between  normalized trajectories. This distance differs from the frequently used \emph{distance correlation}, \cite{Manchein2020,Szkely2007,Mendes2018,Mendes2019} which is a more suitable measure between \emph{cumulative cases}. First, distance correlation is equal to 1 when two sequences have the greatest possible similarity, whereas the distance $D_{ij}$ between two identical sequences is 0. Secondly, whereas sequences $(1,2,3,4)$ and $(4,3,2,1)$ have distance correlation equal to 1, they have significantly different normalized trajectory distance, heralding the fact that one sequence is increasing while the other is decreasing. Specifically, $||\mathbf{g}_i-\mathbf{g}_j||=0$ if and only if $f_i=\alpha f_j$ for some $\alpha>0$, while two sequences $(X_k),(Y_k)$ have distance correlation 1 if and only if $X_k=a+bY_k$ for constants $a,b$. That is, distance correlation does not distinguish between positive and negative gradients.

In Figures \ref{fig:April_Dendrogram}, \ref{fig:May_Dendrogram}, \ref{fig:June_Dendrogram}, \ref{fig:July_Dendrogram}, respectively, we implement hierarchical clustering on these matrices for the months of April, May, June and July. There is consistent similarity in the dendrogram structure: every figure has three clusters, two small clusters, and one majority cluster that contains several subclusters of high internal similarity. The two small clusters generally consist of states that are experiencing steeply increasing or decreasing trajectories, while the larger cluster exhibits more heterogeneity. We describe the common features of the dendrograms in Table \ref{tab:table_distance_months}. There, we also include the \emph{Frobenius norm} of each distance matrix. For an $n \times n$ matrix $A$, this is defined as  $||A||=\left(\sum_{i,j=1}^n |a_{ij}|^2\right)^{\frac{1}{2}}$, and quantifies the total spread of all distances in a month.

\begin{table}[h]
\begin{tabular}{ |p{1.5cm}||p{1.5cm}|p{1.9cm}|p{2.4cm}|}
 \hline
 \multicolumn{4}{|c|}{Trajectory distance matrices} \\
 \hline
 Month & Clusters & Cluster sizes & Frobenius norm \\
 \hline
 April & 3 & \{4,6,41\} & 15.34 \\
 May* & 3 &\{2,2,44\} & 15.97\\
 June & 3 & \{3,14,34\} & 18.35 \\
 July & 3 & \{3,4,44\} & 8.67 \\
\hline
\end{tabular}
\caption{Number of clusters, cluster sizes and Frobenius norm for trajectory distance matrices over four months. *Three states are excluded in May due to low counts.}
\label{tab:table_distance_months}
\end{table}

\begin{figure*}
    \centering
    \begin{subfigure}[b]{0.33\textwidth}
        \includegraphics[width=\textwidth]{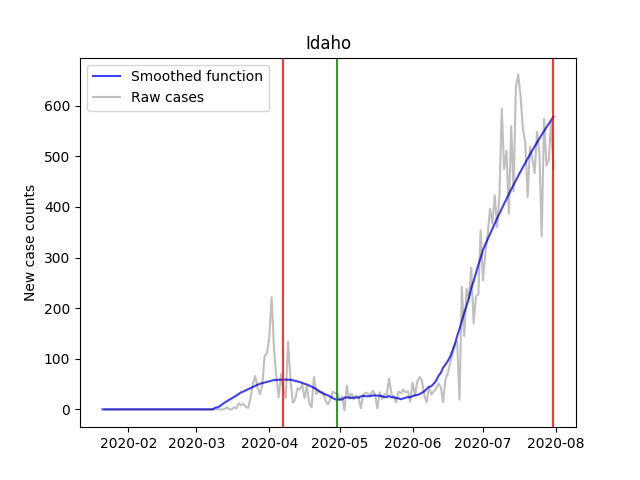}
        \caption{}
        \label{fig:IdahoTS}
    \end{subfigure} 
    \begin{subfigure}[b]{0.33\textwidth}
        \includegraphics[width=\textwidth]{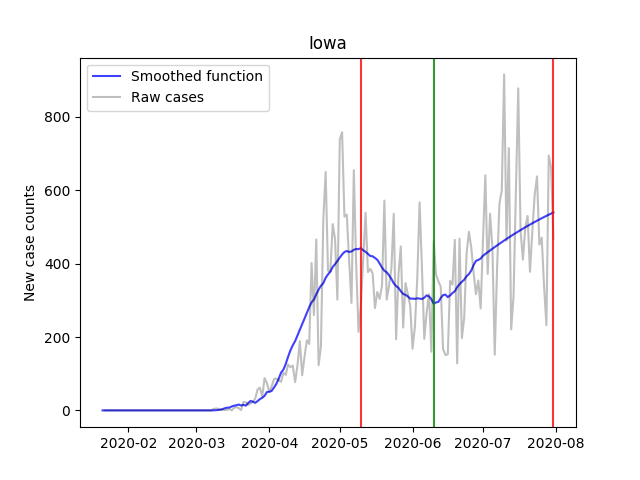}
        \caption{}
        \label{fig:IowaTS}
    \end{subfigure}
    \begin{subfigure}[b]{0.33\textwidth}
        \includegraphics[width=\textwidth]{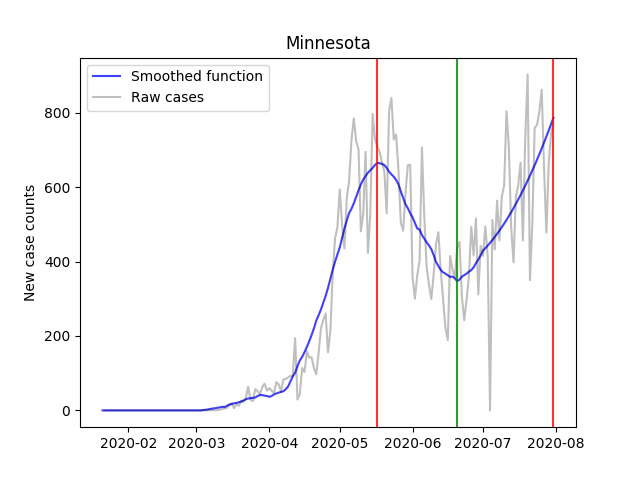}
        \caption{}
        \label{fig:MinnesotaTS}
    \end{subfigure}
    \begin{subfigure}[b]{0.33\textwidth}
        \includegraphics[width=\textwidth]{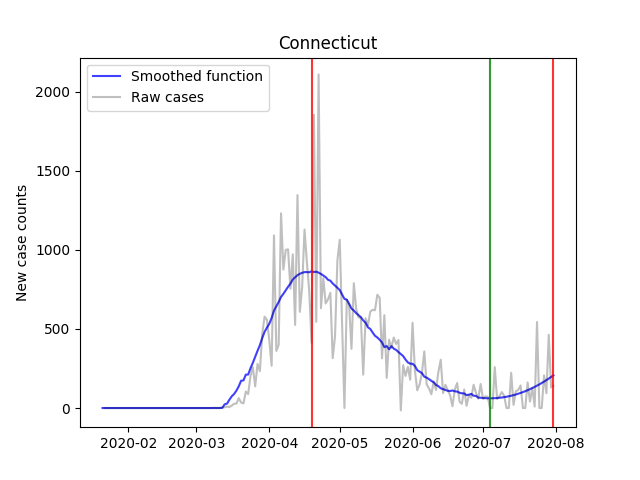}
        \caption{}
        \label{fig:ConnecticutTS}
    \end{subfigure} 
    \begin{subfigure}[b]{0.33\textwidth}
        \includegraphics[width=\textwidth]{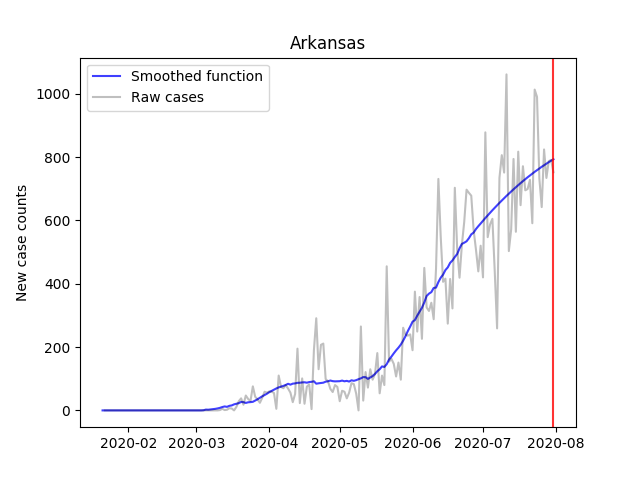}
        \caption{}
        \label{fig:ArkansasTS}
    \end{subfigure}
    \begin{subfigure}[b]{0.33\textwidth}
        \includegraphics[width=\textwidth]{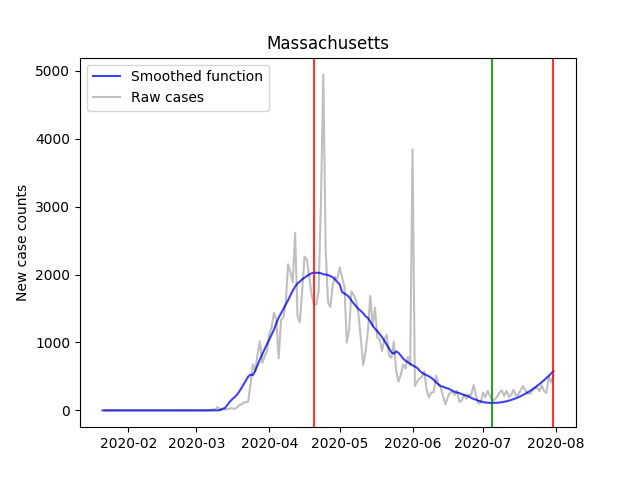}
        \caption{}
        \label{fig:MassachusettsTS}
    \end{subfigure} 
    \caption{Smoothed time series and identified turning points for various states: (a) Idaho, (b) Iowa, (c) Minnesota (d) Connecticut (e) Arkansas, and (f) Massachusetts. Arkansas is in its first surge; Idaho, Iowa and Minnesota are experiencing second surges more severe than the first; Connecticut and Massachusetts are experiencing new increases in cases with peak ratios greater than 0.2, as defined in Section \ref{sec:ChangePointmodel}. Thus, they are determined to be experiencing second surges.}
    \label{fig:section3TSplots}
\end{figure*}

In April, hierarchical clustering determines the existence of three clusters of US states, displayed in Figure \ref{fig:April_Dendrogram}. The first contains Alaska, Hawaii, Idaho, Louisiana, Montana and Vermont, all of which have declining new case trajectories. Idaho, seen in Figure \ref{fig:IdahoTS}, displays behavior typical of the cluster, experiencing a peak in early April, and steadily decreasing for the remainder of the month. The second cluster consists of Iowa, Kansas, Minnesota and Nebraska, all of which have steep increases in new case counts. Iowa and Minnesota are depicted in Figures \ref{fig:IowaTS} and \ref{fig:MinnesotaTS}, respectively. The final cluster contains all 41 remaining states and two subclusters of high self-similarity. The first subcluster contains states whose trajectories are concave down with a local peak in April. Georgia, Pennsylvania and Connecticut, depicted in Figures \ref{fig:GeorgiaTS},  \ref{fig:PennsylvaniaTS} and \ref{fig:ConnecticutTS}, respectively, are typical of this subcluster. The second subcluster consists of states with moderately increasing trajectories, such as Mississippi and  Arizona, depicted in Figures \ref{fig:MississippiTS} and \ref{fig:ArizonaTS}, respectively.

\begin{figure*}
    \centering
    \begin{subfigure}[b]{0.735\textwidth}
        \includegraphics[width=\textwidth]{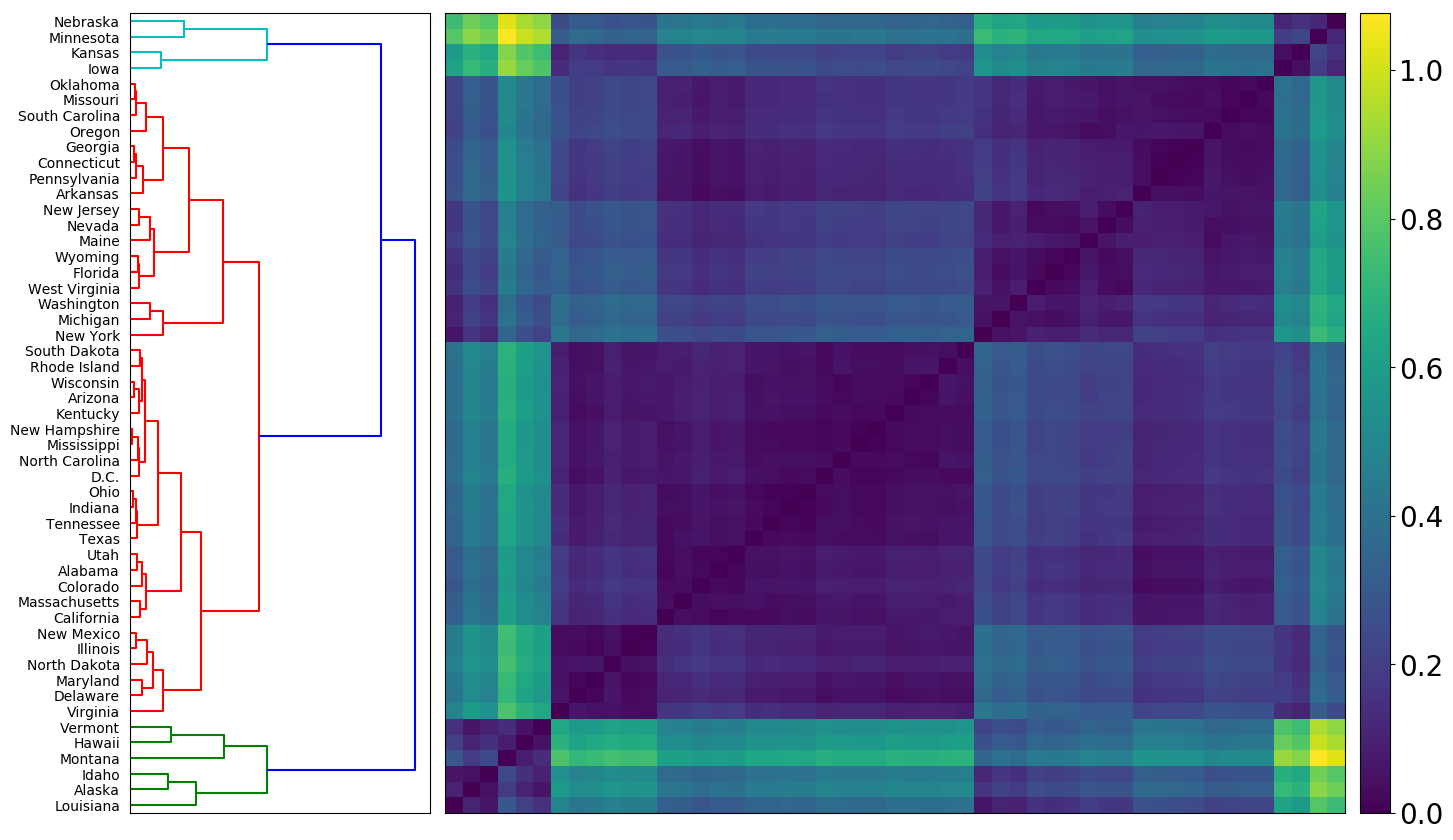}
        \caption{}
        \label{fig:April_Dendrogram}
    \end{subfigure}
    \begin{subfigure}[b]{0.735\textwidth}
        \includegraphics[width=\textwidth]{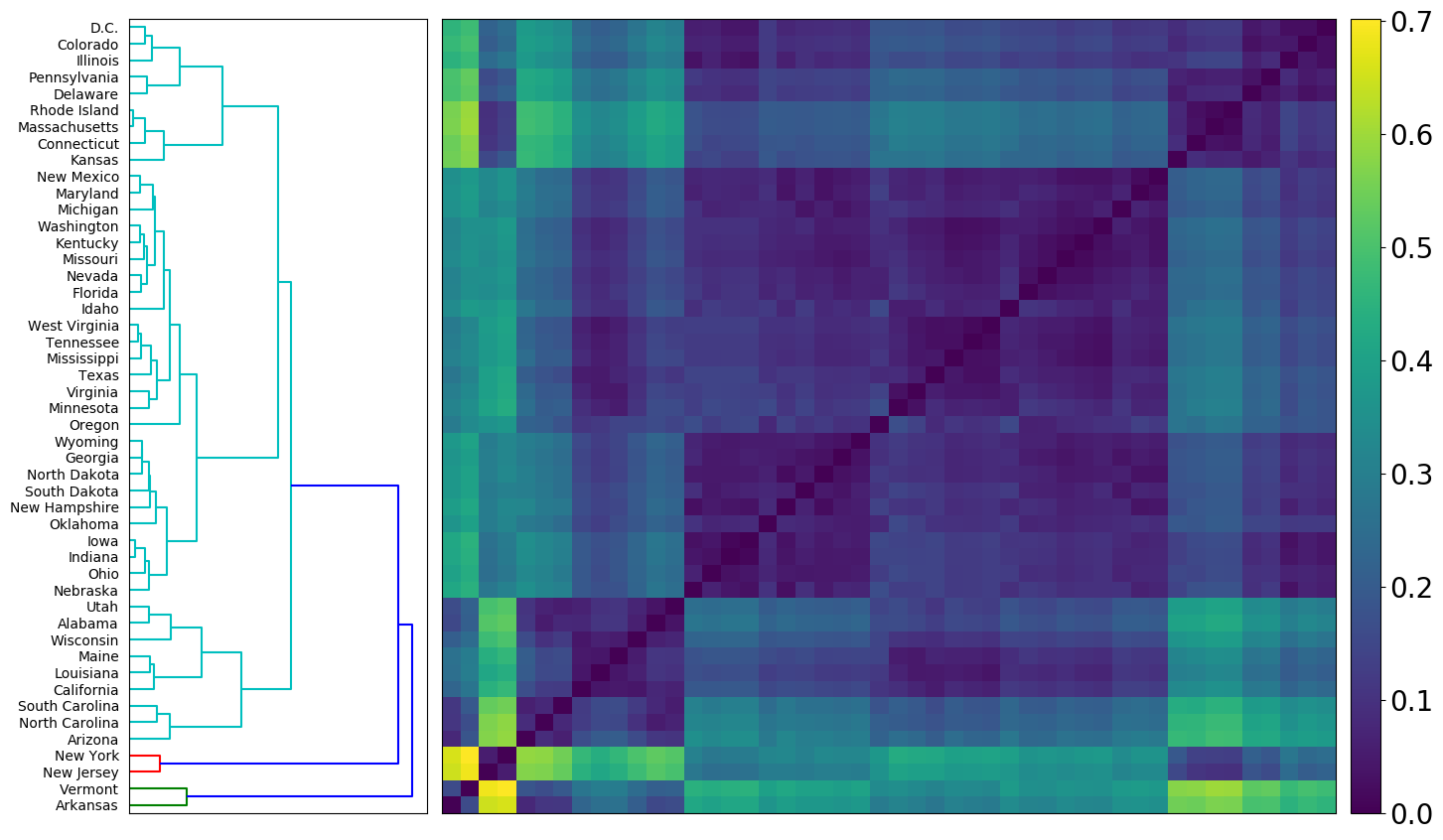}
        \caption{}
        \label{fig:May_Dendrogram}
            \end{subfigure}
        \end{figure*}
        \begin{figure*} \ContinuedFloat
        \begin{subfigure}[b]{0.735\textwidth}
        \includegraphics[width=\textwidth]{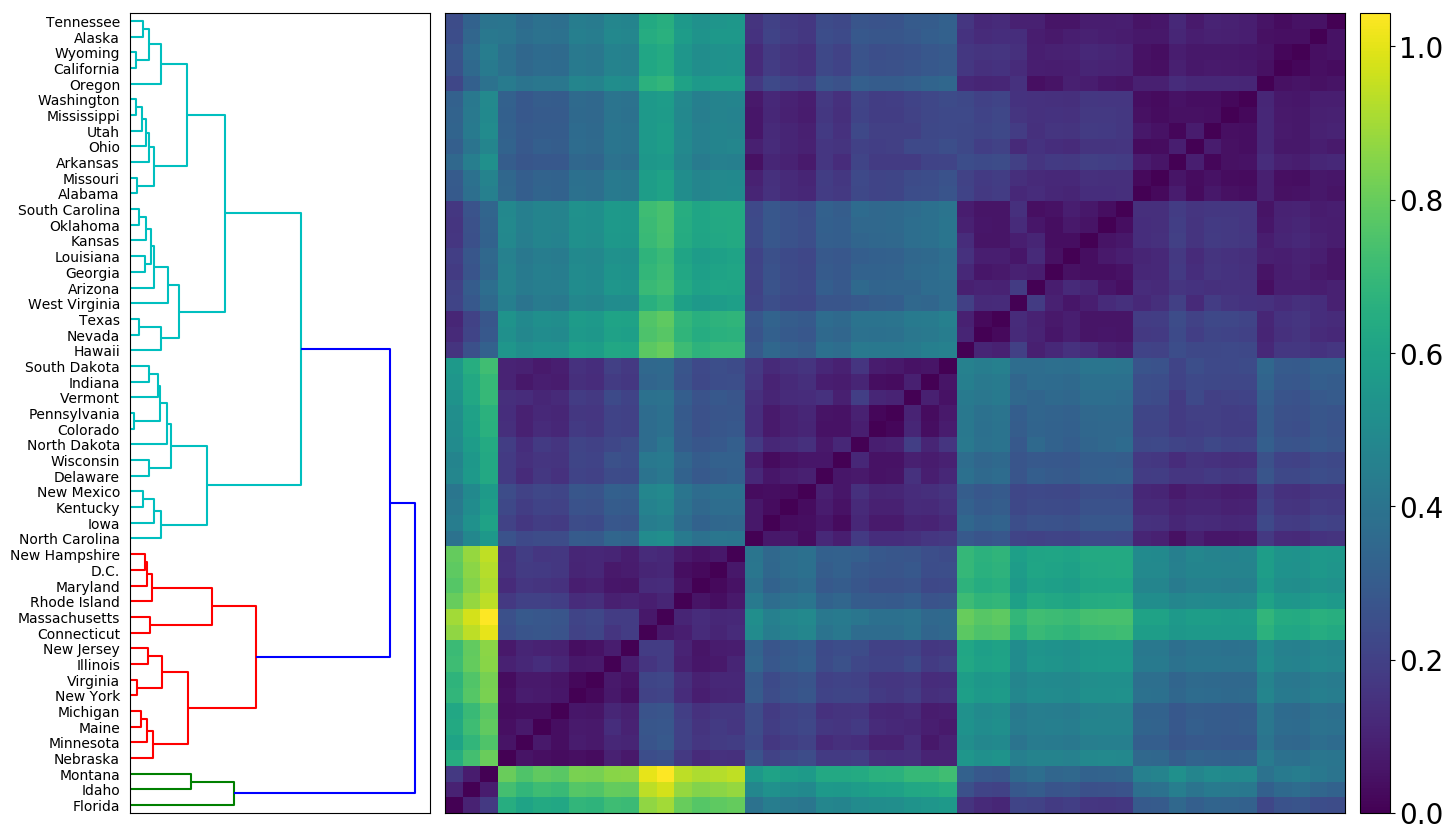}
        \caption{}
        \label{fig:June_Dendrogram}
    \end{subfigure}
    \begin{subfigure}[b]{0.735\textwidth}
        \includegraphics[width=\textwidth]{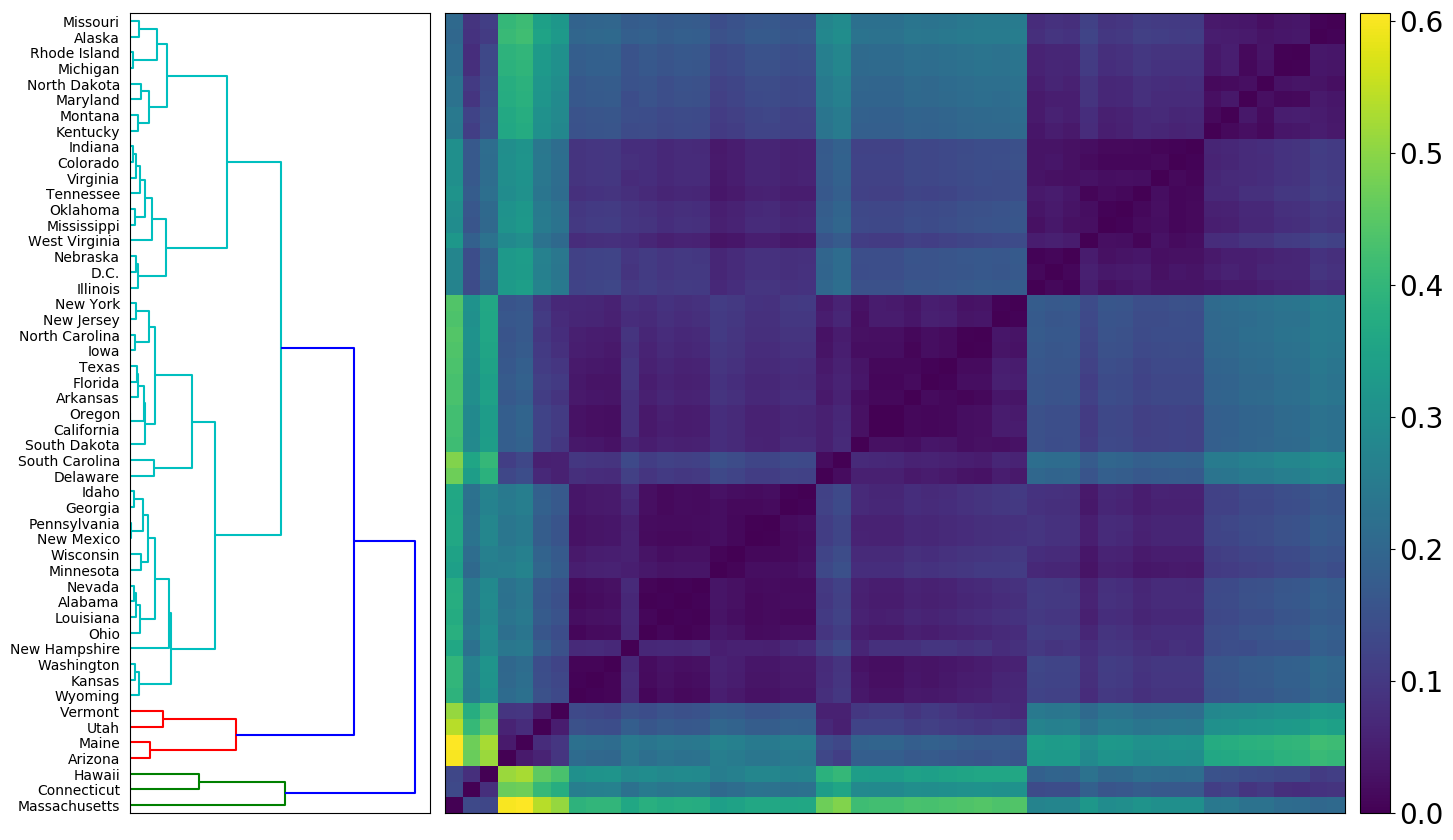}
        \caption{}
        \label{fig:July_Dendrogram}
    \end{subfigure}
    \label{fig:AprilMaydendrograms}
        \caption{\emph{Trajectory dendrograms}, defined in Section \ref{sec:DynamicTrajectoryModelling} for the months of (a) April, (b) May, (c) June and (d) July. These dendrograms highlight states with similar new case trajectories on a month-by-month basis. For all four months there is a consistent cluster structure: 2 small clusters and a large cluster with several concentrated subclusters.}
\end{figure*}


In May, we exclude Alaska, Hawaii, and Montana from the dendrogram (Figure \ref{fig:May_Dendrogram}), as their smoothed trajectories mostly consist of zeroes and very low counts. Again, three clusters are observed: the first and most anomalous cluster contains only New York and New Jersey, whose trajectories are significantly decreasing, as seen in Figures \ref{fig:NewYorkTS} and \ref{fig:NewJerseyTS}, respectively. The second cluster contains Arkansas (Figure \ref{fig:ArkansasTS}) and Vermont; their trajectories are relatively flat at the start of May, with an uptick in the second half. All other states are contained in the final cluster, with several observable subclusters. One notable subcluster contains Northeastern states Connecticut, Delaware, Massachusetts, Pennsylvania, Rhode Island, and D.C. All these states' trajectories are steadily decreasing during May, as seen in Figures  \ref{fig:ConnecticutTS} and \ref{fig:MassachusettsTS} for Connecticut and Massachusetts, respectively. By contrast, another subcluster, containing North Carolina and Arizona (Figures \ref{fig:NorthCarolinaTS} and  \ref{fig:ArizonaTS}, respectively), is characterized by moderate and consistent increase in May.


In June, three clusters are again observed in Figure \ref{fig:June_Dendrogram}. The first consists of Florida, Idaho and Montana, which display significantly increasing new cases. Figures \ref{fig:FloridaTS} and \ref{fig:IdahoTS} show that Florida and Idaho, respectively, experienced high growth from the beginning of June, after a prior month of moderate decrease and flat cases, respectively. In the second cluster, we again observe almost all Northeastern states, including Connecticut (\ref{fig:ConnecticutTS}), D.C., Maine (\ref{fig:MaineTS}), Maryland, Massachusetts (\ref{fig:MassachusettsTS}), New Hampshire, New Jersey (\ref{fig:NewJerseyTS}), New York (\ref{fig:NewYorkTS}), Rhode Island and Virginia. These experienced decreasing trajectories in June from an earlier peak in April. The final cluster is characterized by states with increasing trajectories. A notable subcluster that contains Mississippi, Ohio, Arkansas and others, exhibits considerable similarity in linearly increasing trajectories in June, seen in Figures \ref{fig:MississippiTS},  \ref{fig:OhioTS}, and \ref{fig:ArkansasTS}, respectively.


For July (\ref{fig:July_Dendrogram}), the first of three clusters contains Arizona, Utah, Maine and Vermont. Arizona (\ref{fig:ArizonaTS}) and Utah are experiencing decreases from their first surges, and Maine (\ref{fig:MaineTS}) and Vermont from their second. The second cluster consists of Connecticut (\ref{fig:ConnecticutTS}), Hawaii and Massachusetts (\ref{fig:MassachusettsTS}), all of which exhibit growth in new cases from the beginning to end of July. The states within the final cluster almost all experienced a consistent increase in July, with more nuanced separation of trajectories occurring within the subclusters. For example, one subcluster contains California, Texas and Florida, displayed in Figures \ref{fig:CaliforniaTS}, \ref{fig:TexasTS} and \ref{fig:FloridaTS}, respectively, all of which experience a rapid increase in early July that begins to level off toward the end of July. Even New York and New Jersey experience slight increases in their cases in July, although with much lower absolute counts. Another subcluster contains Georgia, Pennsylvania and Ohio, (\ref{fig:GeorgiaTS}, \ref{fig:PennsylvaniaTS}, \ref{fig:OhioTS}, respectively), which all experience approximately linear growth in new cases. As seen in Table \ref{tab:table_distance_months}, the reduced Frobenius norm for the month of July reflects less spread in the matrix as a whole, due to the large number of states with similarly increasing trajectories.

\section{Conclusion}
\label{sec:conclusion}

In this paper, we propose a new method for analyzing turning points and trajectories among a collection of time series. Our mathematical framework defines the characteristics of a state experiencing (or over) a second surge in COVID-19 cases. The use of semi-metrics between sets of turning points clusters states according to their differing surge behaviors and provides immediate and visible insight into the behavior of the time series collection as a whole.

This classification of behaviors is then accompanied by a close examination of the trajectories on a month-by-month basis. Here we separate and cluster trajectories according to the relative rates of increase or decrease of new case counts. Our methodology is flexible: different smoothing techniques, metrics between data, (semi-)metrics between sets, parameters in the algorithmic framework, and clustering methods can be used to study collections of time series and identify differing surge behavior in greater generality than this application. We demonstrate this with a brief application to the Brazilian federative units in Appendix \ref{Brazil}.


Clustering the states' trajectories on a month-by-month basis reveals consistent similarity in the cluster structure: there are always three clusters, that is, one majority cluster and two smaller clusters. The stable cluster structure over time allows one to easily observe changes in the cluster membership of individual states, and determine the time frame under which new case counts in different states changed direction. For example, in May, New York and New Jersey move into a separate cluster characterized by sharply falling numbers of new cases as they introduced mask mandates.\cite{maskmandateny}


Our analysis provides insights into the evolution of COVID-19 in the United States. While previous papers have studied counts of different countries over shorter time windows, \cite{Machado2020, Manchein2020} this paper studies the US on a state-by-state basis over 7 months. Within our framework, we determine that 31 states plus D.C. are experiencing second surges, of which 21 are more severe than the first surge. 13 states, including 4 of the 10 largest, are still in their first surge, with new case counts that have never materially decreased.  Only 2 states are completely over and 2 partially over their first surge with no second surge as of yet. Just 2 states are over their second. As of the end of July, all other state counts are increasing and 32 exhibited their greatest case counts (after smoothing) on the final day of analysis. All these features are visible in Figure \ref{fig:MJ_Dendrogram}, where five (sub)clusters correspond to these five possible surge behaviors. 

The similarities in Figure \ref{fig:MJ_Dendrogram} can help identify common characteristics of the states that have most and least successfully managed COVID-19. New York and New Jersey, like many of the Northeastern states, experienced peaks in new COVID-19 cases in early April. Unlike other Northeastern states, these three reduced their new cases substantially and have avoided a second surge. Massachusetts and Delaware have experienced small second surges in July, 28.5\% and 55.7\% of their first peak, respectively.

By contrast, California, Texas, Florida and Georgia are four states that managed the growth of new COVID-19 cases poorly: their case counts are the highest in the nation. California and Texas limited restrictions despite cases that never stopped increasing and then reinstated them amid record counts. \cite{California_closing, Texas_masks, Texas_cases} With cases per capita greater than California and Texas, Georgia remains in its first surge, having overturned local mask mandates in July. \cite{georgia_masks} After an early first surge, Florida reduced restrictions and has since experienced a long and steep second surge, including the highest single day counts of any state. \cite{florida_cases} 
Figures \ref{fig:June_Dendrogram} and \ref{fig:July_Dendrogram} place these states in poorly performing clusters, while Figure \ref{fig:MJ_Dendrogram} shows the long second surge of Florida and the continuous first surges of California, Texas and Georgia.

Overall, this paper introduces a new method for analyzing second surge behavior in a collection of time series and provides new insights into the spread of COVID-19 in the US. Early in 2020, many states believed that COVID-19 would resolve quickly. Few predicted that Florida's second surge, for example, would be so much more severe than its first. Nonetheless, this is a highly infectious virus, and even countries that reported zero counts have since observed recurrences. \cite{newzealand} 
As further surges jeopardize both citizen safety and economic recovery, individual states must closely observe the trajectory of their cases and react swiftly to minimize the potential for increasing case counts. We predict that many state governments will learn their lesson from the second surge, and be cautious in observing new case counts. Vigilance going forward is necessary, and we hope states learn this in their response.

\section*{Data availability}
The data that support the findings of this study are openly available at Refs. \onlinecite{datasource} and \onlinecite{brazildata}.
\begin{acknowledgments}
The authors thank Kerry Chen and Orri Ganel for helpful comments and edits.
\end{acknowledgments}

\appendix
\section{COVID-19 in Brazil}
\label{Brazil}
\begin{figure*}
    \centering
    \begin{subfigure}[b]{0.735\textwidth}
        \includegraphics[width=\textwidth]{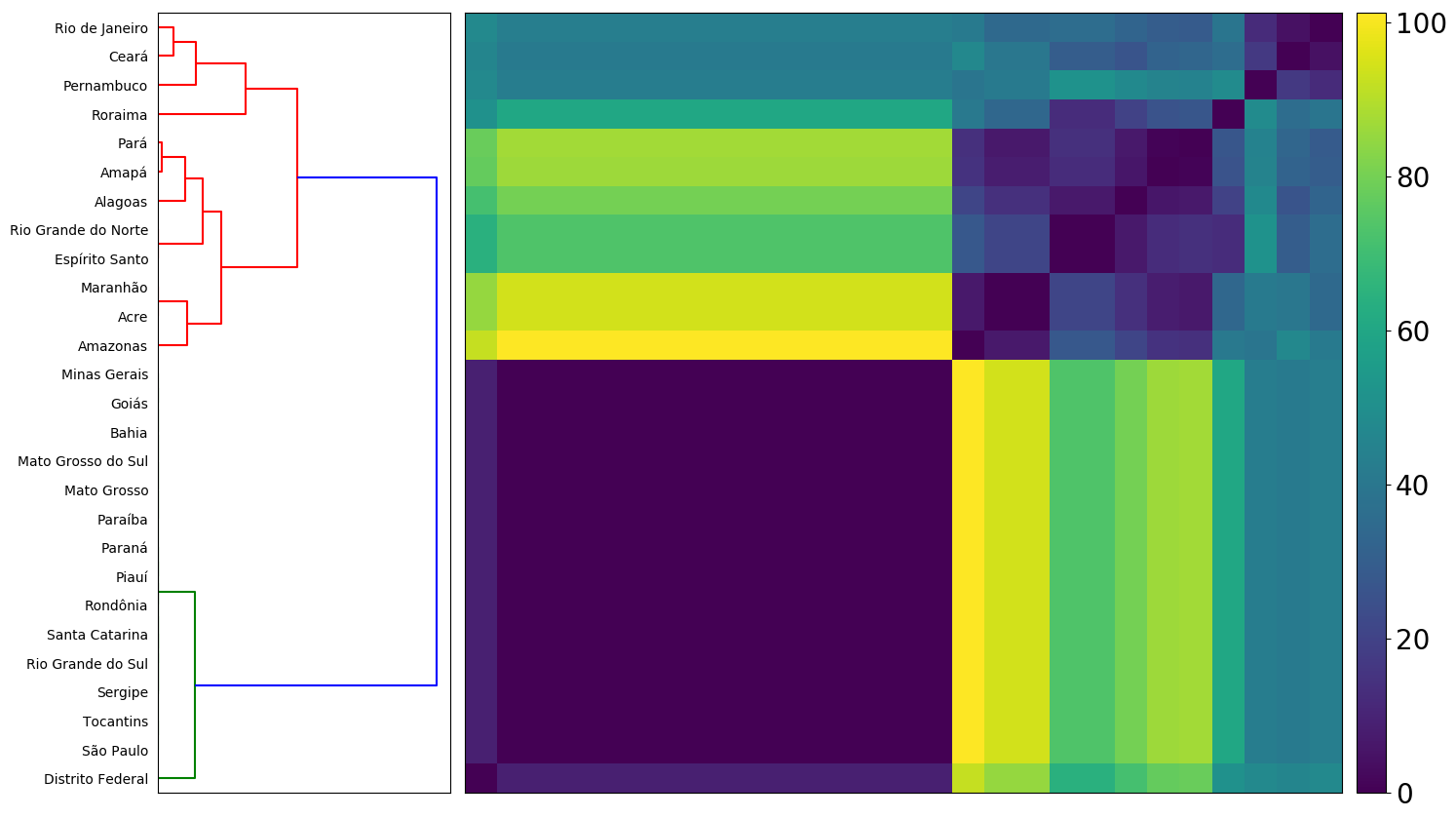}
    \end{subfigure}
    \caption{\emph{Surge behavior matrix}, defined in Section \ref{sec:ChangePointmodel}, applied to the 27 federative units of Brazil. Three clusters of time series are identified with the following behaviors:  14 states plus the federal district in their first surge, 9 states coming down from their first surge, 3 states that are beyond their first surge and are now experiencing a second surge.}
    \label{fig:MJ_BrazilDendrogram}
\end{figure*}

In this brief section, we demonstrate the generality of our method by applying it to the 27 federative units of Brazil. Our data spans 02/25/2020 to 07/31/2020, a period of 158 days. In Figure \ref{fig:MJ_BrazilDendrogram}, we implement hierarchical clustering on the $27 \times 27$ turning point matrix $D^{TP}$ defined in Section \ref{sec:ChangePointmodel}. In contrast with the US, we determine that a majority of 14 states plus the federal district are in their first surge, with new case counts that have never materially decreased. Just 3 states are in their second surge - Cear{\'a}, Pernambuco and Rio de Janeiro, while 9 states are decreasing from their first surge. Unlike US states such as Florida, the second surges are moderate, and are of comparable severity to their first surges. Like the US, we notice significant similarity based on geography. Of the 9 states that are decreasing at the end of the data period, 6 are in the North Region: Acre, Amap{\'a}, Amazonas, Par{\'a}, Rond{\^{o}}nia and Roraima. On the final day of analysis, 15 states exhibited their greatest new case counts (after smoothing).

\section{Algorithmic description of methodology}
\label{algorithm}
In this section, we provide an algorithmic presentation of the computational steps taken for the determination of second surge behavior, described in Section \ref{sec:ChangePointmodel}. Algorithm \ref{alg:stage1} describes the first step of Section \ref{sec:secondsurgemethod}, where an alternating sequence of peaks and troughs is determined. Algorithm \ref{alg:stage2} describes the second step, where this list is refined. In our implementation, we choose three parameters $l=17, \delta=0.2, \epsilon=0.01.$ Equations (\ref{baddefnpeak}) and (\ref{baddefntrough}) define necessary conditions in the first step, while (\ref{loggrad}) defines a necessary condition in the second step.


 \begin{algorithm}[H]
 	\begin{algorithmic} 
 	 	\caption{Turning point identification (step 1)}
 	 \label{alg:stage1}
 	\State Given: a time series $x(t) \in \mathbb{R}$
	\State Form a smoothed time series: $\hat{x}(t)$ = Savitzky-Golay$(x(t))$;
 	\State Data preprocessing: \textbf{If}  {$\hat{x}(t) < 0,  \textbf{ then } \hat{x}(t) = 0$};
	\State Initialize: state = TroughState, Current TP = 1, PeakSet = empty, TroughSet = \{1\};
    \While {Current TP $<T$}
    \State Set $t_0$ = Current TP; Flag = false;
    \For {$t_1=t_0 + 1$ to $T$}
    \If {state = TroughState \textbf{and} $t_1$ satisfies (\ref{baddefnpeak}) \textbf{and} $\hat{x}(t_1)>\hat{x}(t_0)$}
    \State state = PeakState;
    \State Append $t_1$ to PeakSet;
    \State Current TP = $t_1$; Flag = true;
    \State \textbf{Break for}
    
    \ElsIf{state = TroughState \textbf{and} $t_1$ satisfies (\ref{baddefntrough}) \textbf{and} $\hat{x}(t_1)<\hat{x}(t_0)$}
    \State Append $t_1$ to TroughSet;
    \State Remove $t_0$ from TroughSet;
    \State Current TP = $t_1$; Flag = true;
    \State \textbf{Break for}
    \ElsIf{state = PeakState \textbf{and} $t_1$ satisfies (\ref{baddefntrough}) \textbf{and} $\hat{x}(t_1)<\hat{x}(t_0)$}
    \State state = TroughState;
    \State Append $t_1$ to TroughSet;
    \State Current TP = $t_1$; Flag = true;
    \State \textbf{Break for}
     \ElsIf{ state = PeakState \textbf{and} $t_1$ satisfies (\ref{baddefnpeak}) \textbf{and} $\hat{x}(t_1)>\hat{x}(t_0)$}
    \State Append $t_1$ to PeakSet;
    \State Remove $t_0$ from PeakSet;
    \State Current TP = $t_1$; Flag = true;
    \State \textbf{Break for}
    \EndIf
    \EndFor
    \If {Flag = false}
    \State \textbf{Break while}
    \EndIf
    \EndWhile
	\State Output PeakSet and TroughSet.
   \algstore{myalg}
    \end{algorithmic}
    \end{algorithm}
 
  \begin{algorithm}[H]
    \begin{algorithmic}   
    \caption{Turning point refinement (step 2)}
 	 \label{alg:stage2}
    \algrestore{myalg}
  	\State TPSet = Sort(PeakSet $\cup$ TroughSet); \Comment{Indexing begins from 1} 
 	\State Initialize: CurrentPeakIndex = 2; \Comment{Begin the peak ratio refinement}
 	\While{CurrentPeakIndex $\leq$ Length(TPSet) - 2}
 	\State $i$ = CurrentPeakIndex, $t_1$ = TPSet($i$), $t_3$ = TPSet($i+2$);
\If {$\frac{\hat{x}(t_3)}{\hat{x}(t_1)} \geq \delta$}
        \State CurrentPeakIndex = $i+2$;
 \ElsIf {$\frac{\hat{x}(t_3)}{\hat{x}(t_1)} < \delta$ \textbf{and} $i+2$ = Length(TPSet)}
 \State Remove $t_3$ from PeakSet;
\State        TPSet=Sort(PeakSet $\cup$ TroughSet);
\ElsIf{ $\frac{\hat{x}(t_3)}{\hat{x}(t_1)} < \delta$ \textbf{and} $i+2 <$ Length(TPSet)}
\State $t_2$ = TPSet($i+1$), $t_4$ = TPSet($i+3$);
    \If {$\hat{x}(t_2) \leq \hat{x}(t_4)$} 
\State Remove $t_4$ from TroughSet;
\Else 
\State Remove $t_2$ from TroughSet;
\EndIf
\State Remove $t_3$ from PeakSet;
\State        TPSet = Sort(PeakSet $\cup$ TroughSet);
\EndIf
 	\EndWhile

\State Initialize: CurrentIndex = 1; \Comment{Begin the log-grad refinement}
\While{CurrentIndex $<$ Length(TPSet)}
\State $i$ = CurrentIndex, $t_0$ = TPSet($i$), $t_1$ = TPSet($i+1$);
\If{$|\loggrad(t_0, t_1)|<\epsilon$} \Comment{See Equation (\ref{loggrad})}
\State Remove $t_0$ and $t_1$ from both TroughSet and PeakSet;
\State TPSet = Sort(PeakSet $\cup$ TroughSet);
\Else 
\State CurrentIndex = $i+1$;
\EndIf
\EndWhile
\State Output PeakSet and TroughSet.
 	
    \end{algorithmic}
    \end{algorithm}
 
\bibliography{aipsamp}

\end{document}